\documentclass[11pt,a4paper]{article}

\usepackage{natbib}

\newif\ifpdf 
\ifx\pdfoutput\undefined 
\pdffalse 
\else 
\pdfoutput=1 
\pdftrue 
\fi

\ifpdf
\usepackage{ae}
\usepackage[pdftex]{graphicx}
\else
\usepackage{graphicx}
\fi

\usepackage{amssymb}
\usepackage{latexsym}

\newcommand{\fig}[1]{Fig.~\ref{#1}}

\newcommand{\eq}[1]{Eq.~\ref{#1}}
\newcommand{\eqs}[1]{Eqs.~\ref{#1}}

\renewcommand{\d}{{\rm d}}

\newcommand{\ub}{{\rm ub}}
\newcommand{\bd}{{\rm b}}
\newcommand{\piad}{\pi_{\rm ad}}
\newcommand{\tpi}{\tilde\pi_{\rm ad}}

\begin{document}

\begin{center}
  {\huge Self-organized density patterns of molecular motors in arrays
    of cytoskeletal filaments}\\[6mm]
  {\large Stefan Klumpp,$^*$ Theo M.\ Nieuwenhuizen,$^{*,\#}$ and
    Reinhard
    Lipowsky$^*$}\\[5mm]
\end{center}

\noindent
$^*$ Max-Planck-Institut f\"ur Kolloid- und
Grenzfl\"achenforschung, 14424 Potsdam, Germany, and $^\#$ Instituut voor Theoretische Fysica, Valckenierstraat 65, 1018 XE Amsterdam, The Netherlands\\[3mm]

\noindent
Keywords: Active processes, molecular motors, kinesin, traffic jams,
exclusion, pattern formation

\noindent
Running title: Density patterns of molecular motors\\[3mm]

\noindent
Version: \today \\[3mm]

\begin{abstract}
  The stationary states of systems with many molecular motors are
  studied theoretically for uniaxial and centered (aster-like)
  arrangements of cytoskeletal filaments using Monte Carlo simulations
  and a two-state model. Mutual exclusion of motors from binding sites
  of the filaments is taken into account.  For small overall motor
  concentration, the density profiles are exponential and algebraic in
  uniaxial and centered filament systems, respectively. For
  uniaxial systems, exclusion leads to the coexistence of regions
  of high and low densities of bound motors corresponding to motor traffic
  jams, which grow upon increasing the overall motor concentration.
  These jams are insensitive to the motor behavior at the end of the filament. 
  In centered systems, traffic jams remain small and an increase
  in the motor concentration leads to a flattening of the profile, if
  the motors move inwards, and to the build-up of a concentration
  maximum in the center of the aster if motors move outwards. In addition 
  to motors density patterns, we also determine the corresponding patterns 
  of the motor current. 
\end{abstract}
\newpage

\section*{Introduction}

Cytoskeletal motors such as kinesin, dynein, and myosin are proteins
which convert the chemical free energy released from the hydrolysis of
adenosine triphosphate (ATP) into directed movements along filaments
of the cytoskeleton. In cells, these motors drive various transport
processes, and are also involved in cell division, cell locomotion,
and reorganization of the cytoskeleton
\citep{Schliwa_Woehlke2003,Howard2001}.  A lot of knowledge has been
obtained from in vitro motility assays which allow for the measurement
of single motor properties such as their velocities, average walking
distances, step sizes, and the forces they exert \citep{Howard2001}.
These quantities have been measured for various types of processive
motors including conventional kinesin
\citep{Howard__Vale1989,Block__Schnapp1990,Svoboda__Block1993,Meyhofer_Howard1995,Vale__Yanagida1996,Schnitzer_Block1997},
Myosin V \citep{Mehta__Cheney1999,Veigel__Molloy2002}, the processive
monomeric kinesin KIF1A
\citep{Okada_Hirokawa1999,Tomishige__Vale2002}, and cytoplasmic dynein
\citep{Wang_Sheetz2000,King_Schroer2000}.  These motility assays study
systems consisting either of (i) mobile motors and immobilized
filaments or of (ii) immobilized motors and mobile filaments. In
addition, systems where (iii) both motors and filaments are mobile and
filaments can be displaced by motors have also be studied \citep[see
e.g.][]{Takiguchi1991,Urrutia__Kachar1991,Nedelec__Leibler1997,Surrey__Karsenti2001,Kruse_Juelicher2000}.

In all of these systems, motors and filaments interact via  hard core
interactions arising from their mutual exclusion. Indeed, 
both motors and  filaments
occupy a certain spatial volume which cannot be occupied by another
molecular structure.  In particular, motors
bound to filaments exclude other motors from the binding sites of the  
filaments.  The latter exlusion effects were first addressed in our 
previous work  \citep{Lipowsky__Nieuwenhuizen2001} in which we introduced a 
general class of driven lattice gas models for this purpose. 

In the following, we use these driven lattice gas models in order
to explore how the arrangement of the filaments affects the 
motor transport in closed compartments. We consider {\em uniaxial} and 
{\em centered}  filament arrangements and present results for the stationary
patterns of both motor density and motor current.  Both types of 
arrangements are accessible to in vitro experiments
and mimic structures
of the cytoskeleton as observed in vivo. The uniaxial systems mimic the
geometry of axons or fungal hyphae, while centered systems are realized, for
example, in the aster-like structures of microtubules extending from
centrosomes.
For the uniaxial systems, we have previously shown that traffic jams
build up easily as a consequence of mutual exclusion
\citep{Lipowsky__Nieuwenhuizen2001}, while previous work on centered
systems \citep{Nedelec__Maggs2001} did not incorporate this mutual
exclusion.

We will show in the following that uniaxial and centered systems
exhibit rather different jamming behavior.  While in uniaxial systems
jammed regions grow upon increasing the motor concentration and spread
over the whole system, the effect of jamming in centered systems is 
less dramatic and
jams remain small in this case. Increasing the motor concentration,
however, influences the density profile in the non-jammed region.  In
addition, we show that the traffic jams in uniaxial systems are rather
insensitive to the motor behavior at the end of the filaments. In
contrast, the latter behavior is crucial for the presence of jams in
centered systems.

The density profiles discussed here theoretically can be directly
measured in biomimetic experiments in vitro, and, in fact, such
density profiles have recently been measured for the case of centered
or aster-like systems \citep{Nedelec__Maggs2001}. However, the latter
experiment did not address the jamming behavior, which could be done by
increasing the motor concentration in these systems. In addition, our
theoretical density profiles can be compared to motor density profiles
measured for the corresponding systems in vivo. Such in vivo density
profiles have been reported for fungal hyphae, which represent
uniaxial systems. \citet{Seiler__Schliwa2000} have observed motors
localized at the tip of these hyphae, which corresponds again to the
case of low motor density.  In vivo, the motor concentration can be
changed by changing the level of expression of the corresponding gene;
in that way jam-like density profiles have recently been observed for
another fungal kinesin-like motor \citep{Konzack2004,Konzack__Fischer2004}.  
The effect of exclusion (and, thus, jamming) is
enhanced if the motors transport large cargoes such as membranous
organelles. Jam-like behavior of organelles has been observed in axons (W.
Saxton, private communication), extreme cases induced by mutations of
motors (which are lethal in later stages of development) are accompanied 
by strong swelling of the axon
\citep{Hurd_Saxton1996,Martin__Saxton1999}.

Our article is organized as follows. We introduce the theoretical
model in the following section.
In the sections 'Density profiles for uniaxial filament systems' and 'Density 
profiles for centered filament systems' 
we discuss jamming effects in two types of filament systems and present results for the motor density patterns obtained from Monte Carlo simulations and from
a two-state model. 
Finally, we relate our
results to recent experiments
in the discussion section. 
The appendices describe the theoretical methods 
used in this article and some analytical calculations.


\section*{Lattice models for molecular motors and filaments}
\label{sec:model}

\begin{figure}[h]
  \begin{center}
    \leavevmode
    \includegraphics[angle=0,width=.8\textwidth]{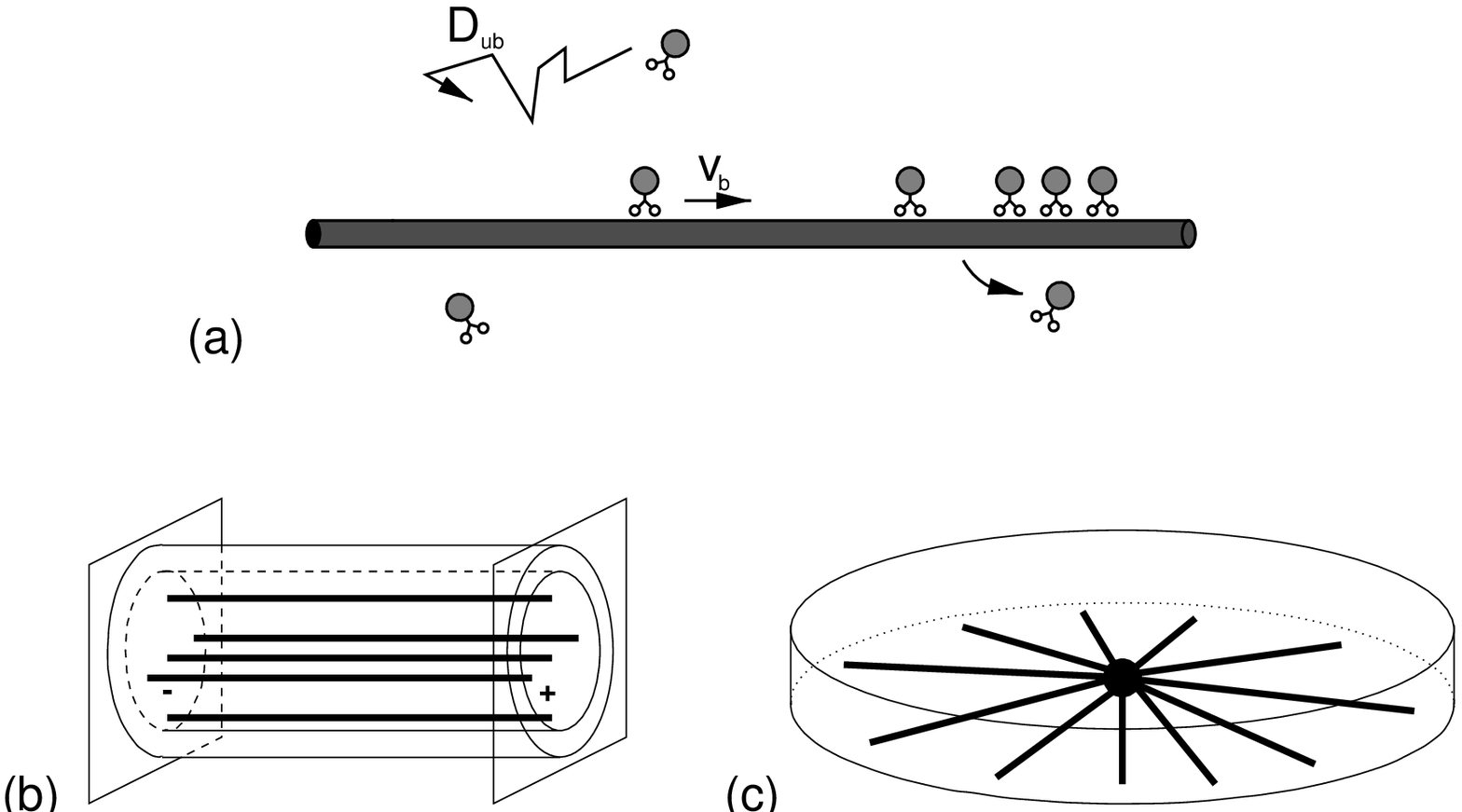}
    \vspace{5cm}\caption{}
    \label{fig:tubeAsterGeom}
  \end{center}
\end{figure}

In this article, we study the stationary profiles of the motor density
which build up within closed compartments containing filaments.  These
stationary states are characterized by the balance of bound and
unbound motor currents \citep{Lipowsky__Nieuwenhuizen2001}. Unbinding
of motors from the filaments reflects the finite binding energy of the
motor--filament complex which can be overcome by thermal fluctuations
and leads to peculiar random walks of the motors which consist of
alternating sequences of directed motion along filaments and
non-directed diffusion in the surrounding fluid
\citep{Ajdari1995,Lipowsky__Nieuwenhuizen2001,Nieuwenhuizen__Lipowsky2002,Nieuwenhuizen__Lipowsky2004},
see \fig{fig:tubeAsterGeom}(a).  In order to study these random walks,
we have recently introduced lattice models
\citep{Lipowsky__Nieuwenhuizen2001}. One useful feature of these
models is that one can incorporate motor--motor interactions such as
the mutual exclusion in a rather natural way
\citep{Lipowsky__Nieuwenhuizen2001,Klumpp_Lipowsky2003,Klumpp_Lipowsky2004}.
Motor--motor interactions are especially important on the filaments:
motors are strongly attracted to filaments, so that the local density
of motors on these filaments will typically be large even if the
overall motor concentration is rather small. The importance of
motor--motor interactions is further increased if motors accumulate in
certain regions of closed compartments.

Mutual exclusion of motors from binding sites of the filaments has two
effects: (i) Binding of motors to the filament is reduced for those
filament segments which are already occupied by many motors. This
effect is directly observed
in decoration experiments 
\citep[see, e.g.,][]{Song_Mandelkow1993,Harrison__Johnson1993}.  (ii)
The mutual hindrance slows down the movement of motors in regions of
high motor density.  This second effect has not yet been studied
experimentally, but there are indications of it in microtubule gliding
assays \citep{Boehm__Unger2000a}.  In addition, there is indirect
evidence for such a slowing down from the self-organization of
microtubules and motors, where an increase of motor concentration can
induce a transition from vortex to aster patterns of microtubules
\citep{Surrey__Karsenti2001}.  From computer simulations, such a
transition is expected if the motors spend more time close to the end
of a filament. This should happen if the motors are slowed down at the
filament end by a traffic jam which builds up upon increasing the
motor concentrations.

\subsection*{Bound and unbound motor movements}

In the following, we describe the movements of molecular motors as
random walks on a three-dimensional cubic lattice
\citep{Lipowsky__Nieuwenhuizen2001,Nieuwenhuizen__Lipowsky2002,Nieuwenhuizen__Lipowsky2004}.
One or several lines of lattice sites represent one or several
filaments.  The lattice constant $\ell$ is taken to be the repeat
distance of the filament which is 8~nm for kinesins moving along
microtubules, so that filament sites of the lattice correspond to
binding sites of the filament. A motor bound to a filament performs a
biased random walk which describes the active movements along the
filament.  Per unit time $\tau$, it attempts to make forward and
backward steps with probability $\alpha$ and $\beta$, respectively. As
backward steps are rare for cytoskeletal motors, we take $\beta=0$ in
the following which eliminates one parameter from our systems. Rather
similar behavior is found for small nonzero values of $\beta$. With
probability $\gamma$, the bound motor makes no step, and with
probability $\epsilon/6$, it unbinds to each of the four adjacent
non-filament sites.  The sum of all hopping probabilities per unit
time $\tau$ is one, i.e.\ the probabilities are related by
\begin{equation}\label{sumProb}
  \alpha+\beta+\gamma+4\epsilon/6=1.
\end{equation}

When the motor particle reaches the end of the filament, it does not
have the possibility to step forward to another filament site.  We
will consider two different unbinding processes for this `last'
filament site as in our previous work (Klumpp and Lipowsky, 2003):

-- {\em thermal unbinding:} the motor particle detaches from the
`last' filament site with probability $\epsilon/6$ to the unbound site
in the forward direction, but remains at the 'last' site with
probability $\gamma'\equiv\gamma+\alpha-\epsilon/6$, while the
backward probability $\beta$ and the sideward probability $\epsilon/6$
remain unchanged. Adjusting the no-step probability implies the
modified normalization
\begin{equation}\label{thermalunbindlast}
  \beta + \gamma' +5 \epsilon/6 = 1
\end{equation}
for the hopping probabilities at the `last' filament site.

-- {\em active unbinding:} the motor particle detaches from the `last'
site with probability $\alpha$ in the forward direction and with
probability $\epsilon/6$ in the four sideward directions as for all
other filament sites. In this case, the normalization of the hopping
probabilities at the `last' filament site is given by (1).

An unbound motor performs a symmetric random walk which corresponds to
non-directed diffusive movement. It attempts to step to each adjacent
lattice site with equal probability $1/6$. If an unbound motor reaches
a filament site, it can bind to this site with probability $\piad$.
The random walk probabilities can be chosen in such a way that one
recovers the measured transport properties of specific motors such as
the bound state velocity, the unbound diffusion coefficient and the
average walking distance 
\citep[see][]{Lipowsky__Nieuwenhuizen2001,Klumpp_Lipowsky2003}.\footnote{Note
  that the model used here does not account for the bound state
  diffusion coefficient or, equivalently, the randomness parameter of
  the motor movements. This parameter can be incorporated by
  introducing a second time scale for the movements of the bound
  motors \citep[see][]{Lipowsky__Nieuwenhuizen2001}.  Such an
  externded model leads to density profiles which are very similar to
  the ones described here. This indicates that the overall diffusion
  is essentially governed by the unbound diffusion process.} The
unbound diffusion coefficient $D_\ub$ fixes the basic time scale
$\tau=\ell^2/D_\ub$. The probabilities $\alpha$, $\beta$, $\gamma$,
and $\epsilon$ are determined from the velocity
$v_\bd=(\alpha-\beta)\ell/\tau$ of a single bound motor, the average
walking distance along the filament $\Delta
x_\bd=3v_\bd\tau/(2\epsilon)$, the condition $\beta=0$ and
\eq{sumProb}.

Mutual exclusion of motors is taken into account by rejecting all
hopping attempts to lattice sites which are occupied by other motors.
We take the motor particles to have a linear size comparable to the
filament repeat distance $\ell$ and to occupy a volume $\ell^3$. If
the motors are attached to larger cargoes, exclusion is enhanced.  In
particular, a large cargo of linear size $n\ell$, when bound to the
filament, effectively covers between $n\ell$ and $(2n-1)\ell$ filament
sites depending on the bound density. However, the functional
relationships between the different densities and current are rather
similar \citep{MacDonald__Pipkin1968,McGhee_vonHippel1974}. We will
briefly discuss this case at the end of the paper in the 'Discussion'
section.

These lattice models for systems with many molecular motors are
related to driven lattice gas models which have been studied
extensively in the context of non-equilibrium phase transitions
\citep{Katz__Spohn1983,Krug1991}. In the models studied here, the
'driving', i.e.\ the active directed movement, is restricted to the
linear subspaces corresponding to the filaments.

In the following, we will usually express all lengths and times in
units of the filament repeat distance $\ell$ and the basic time scale
$\tau$, respectively. This means that the bound and unbound motor
densities $\rho_\bd$ and $\rho_\ub$ that we will consider in the
following are local particle number densities satisfying $0\leq
\rho_\bd\leq 1$ and $0\leq \rho_\ub\leq 1$, which corresponds to
$0\leq \rho_\bd\leq 1/\ell^3$ and $0\leq \rho_\ub\leq 1/\ell^3$ in
dimensionful units. Dimensionful units will be used when presenting
results for specific motor molecules.

\subsection*{Filament arrangements and compartment geometries}

In this article, we study two types of filament arrangements within 
closed compartments as
shown in \fig{fig:tubeAsterGeom}(b) and (c). The first type is a uniaxial 
filament system where a closed cylindrical tube contains a number $N_f$
of uniaxially arranged filaments, i.e.\ filaments oriented parallel to
the cylinder axis and with the same orientation.
We denote the coordinate parallel to the filament by $x$ and the
coordinates perpendicular to it by $y$ and $z$. The tube has length
$L$ and radius $R$.

The second type of system which we will study is a centered filament 
system, i.e., 
a radial or aster-like arrangement of
filaments within a closed disk-like compartment. The number of
filaments is again $N_f$. In this case, we denote the radial
coordinate by $r$. The linear extension of the compartment along the
direction of the filaments, i.e.\ the disk radius, is denoted by $L$
and the disk height by $h$.

In both cases, we take all filaments to have the same length which
is equal to the corresponding linear extension of the compartment,
i.e.\ to the tube length and to the disk radius in the case of
uniaxial and centered filament systems, respectively. Shorter
filaments lead to very similar results. An example would be filaments
in radial arrangements which are nucleated from a centrosome and
extend from $r=R_c$, the centrosome radius, to the disk radius $r=R$.
In addition, since the compartments are closed, and the number of motors,
denoted by $N$, stays constant within each compartment.


\section*{Density profiles for uniaxial filament systems}
\label{sec:tube}

We first consider uniaxial arrangements of filaments within a closed
tube as shown in \fig{fig:tubeAsterGeom}(b). On the one hand, placing
one or several filaments and motors inside a tube should be
experimentally feasible. The tube could be either a glass tube as used
for micropipettes, a topographic channel as used for filament guiding
\citep{Clemmens__Vogel2003}, or a liquid microchannel on a chemically
structured surface \citep{Gau__Lipowsky1999,Brinkmann_Lipowsky2002}.
In all cases, tube diameters down to a few $\mu$m can be achieved. On
the other hand, tube-like geometries are also quite common in cells,
the most prominent example being the axon of a nerve cell, a tubular
cell compartment with a diameter in the range of few micrometers 
and a length of up to a meter, which contains tens of microtubules per 
$\mu{\rm m}^2$ \citep{Alberts__Walter2002}, i.e., typical distances of 
the microtubules are in the range of 100~nm.  Similar compartments,
the hyphae, exist in the case of fungal cells. In addition, some
compartments inside the cell have tubular shapes and contain filaments 
such as strands of
cytosol crossing vacuoles in plant cells, again with diameters in the 
micron range.

We will now focus on the case of a single filament, since the case of
$N_f$ isopolar parallel filaments in a tube with cross-section $\phi$
is essentially equivalent to a single filament in a tube with
cross-section $\phi/N_f$ provided that the filaments are equally
distributed within the tube. (If the filaments are concentrated in a
certain region, i.e., if the distance between filaments is small
compared to the distance between filaments and the tube wall,
depletion of motors is enhanced; depletion effects are rather weak,
however.) Let us consider a cylindrical tube of length $L$ and radius
$R$ with one filament located along its symmetry axis.  Imagine now
that a certain number of motors is placed into this tube.  In the
absence of ATP, the system attains an equilibrium state, where binding
to and unbinding from the filament balance each other locally, i.e.\ 
at every single binding site. Both the bound and the unbound motor
densities are constant and related by the radial equilibrium condition
\begin{equation}\label{equilibrium_withExclusion}
  \pi_{\rm ad}\rho_\ub (1-\rho_\bd)=\epsilon\rho_\bd(1-\rho_\ub)\approx\epsilon\rho_\bd, 
\end{equation}
where the terms $(1-\rho_\bd)$ and $(1-\rho_\ub)$ describe mutual
exclusion of bound and unbound motors, respectively, with
$(1-\rho_\ub)\approx 1$ for typical experimental situations.

When ATP is added to the system, the motors start to move along the
filament.  We use the convention that the filaments are oriented in
such a way that the bound motors move to the right.  The motor current
along the filament builds up a density gradient, which generates a
diffusive current. In the stationary state, this diffusive current
balances the drift current of bound motors.  As a first approximation,
we assume that \eq{equilibrium_withExclusion} is also valid in the
presence of ATP (which is justified if the velocity $v_\bd$ is
sufficiently small, as we will show below).  The balance of currents
can then be expressed by
\begin{equation}\label{firstApprox}
  v_\bd\rho_\bd (1-\rho_\bd)=D_\ub\phi\,\frac{\partial\rho_\ub}{\partial x}\simeq D_\ub\phi\frac{\epsilon}{\pi_{\rm ad}}\,\frac{\partial}{\partial x}\frac{\rho_\bd}{1-\rho_\bd}.
\end{equation}
It follows from this relation that, for low motor densities, the motor
density increases exponentially along the filament and that motors
accumulate at the right end of the filament, further increasing the
importance of exclusion effects there.

\subsection*{Simulation results}

Typical density profiles as obtained from Monte Carlo simulations are
shown in \fig{fig:dichteProfileTube} (a). If the total number $N$ of
motors is relatively small --- one example is provided by $N=100$ in
\fig{fig:dichteProfileTube} --- motors are essentially localized at
the right end of the tube. Crowding of motors occurs only in a short
region at the end of the filament where motors form a kind of traffic
jam. To the left of the traffic jam, the density has an exponential
profile as predicted by the simple balance of active directed currents
and diffusive currents.  If a motor detaches from the filament in the
crowded region, it will diffuse back over a certain distance and most
likely rebind to the filament in the region to the left of the traffic
jam. In the jammed region rebinding is strongly reduced, since
essentially all binding sites are already occupied. Upon rebinding to
the filament, the motor will move relatively fast to the right until
it ends up in the jammed region again.

These observations imply the coexistence of a low-density region with
an exponential density profile and a crowded high-density region,
separated by a relatively sharp domain wall or interface which
corresponds to the beginning of the traffic jam.  If the number of
motors in the tube is increased, the jammed region spreads further to
the left and the domain boundary is shifted towards smaller values of
the spatial coordinate $x$, as shown in \fig{fig:dichteProfileTube}
for $N=350$.  Now, motors diffuse backwards over larger distances,
since attempts to rebind to the filament fail, if the binding sites
are already occupied.

Finally, if the motor concentration is very large, there is only one
domain with a high density of bound motors: The filament is crowded
over its whole length and the bound density profile is essentially
constant except for the regions close to the two ends of the filament,
see the case $N=1000$ in \fig{fig:dichteProfileTube}. In this case,
motors may diffuse back over the whole system length, but both the
diffusive current and the bound current along the filament are very
small.

The corresponding profiles of the bound motor current along the
filament are shown in \fig{fig:dichteProfileTube}(b). The diffusive
current of unbound motors integrated over the tube cross-section has
the same absolute value, but the opposite sign.  The current depends
strongly on the position $x$ along the filament as long as the
filament is not completely jammed. Like the motor densities, it
increases exponentially in the low density region. In the jammed
region at the right end of the tube, the current decreases rapidly. It
reaches its maximum close to the end of the traffic jam.  Note that
the maximal current accessible in these systems is smaller than
$v_\bd/4$, i.e.\ smaller than the maximally possible current in a
system with constant densities such as a tube system with periodic
boundary conditions \citep{Klumpp_Lipowsky2003}. For the case of the
completely jammed filament, the current profile is nearly flat, while
the absolute current is small.

To obtain a global characterization of transport in the system, we
determined the average current defined by
\begin{equation}
  \bar J_\bd \equiv \frac{1}{L}\int_0^{L} \d x \, j_\bd(x).
\end{equation}
This quantity exhibits a maximum at an optimal motor concentration as
a function of the total number $N$ of motors in the system, i.e., as a
function of the overall motor concentration, see \fig{fig:stromMax}.
For small $N$, it grows linearly with the number of motors, while for
large $N$ it decreases again since motion of the bound motors is
slowed down by the increasing traffic jam.  However, this decrease of
the currents is rather slow, since additional motors introduced into
the system can only rarely find free binding sites. For the system
shown in \fig{fig:stromMax}, the maximal current occurs for $N\simeq
350$ motors, which corresponds to the intermediate case of the
profiles in \fig{fig:dichteProfileTube}.

A second quantity, which gives a global characterization of the
profiles, is the traffic jam length $L_*$ of the crowded domain.
$L_*$ can be defined by the condition $\rho_\bd(x_*)=1/2$ via
$L_*=L-x_*$.  Results for $L_*$ are also shown in \fig{fig:stromMax}.
The three cases discussed above can now be distinguished as follows.
For very small $L_*/L$, crowding of motors only occurs in a small
region at the filament end and the profile decays exponentially to the
left over a large fraction of the system size. For intermediate values
of $L_*/L$ with $0\ll L_*/L \ll 1$, the density profiles exhibit
coexistence of domains with high and low bound motor densities.
Finally, for $L_*/L\approx 1$ the whole filament is crowded. Comparing
the functional dependence of the traffic jam length $L_*$ with the
average bound current $\bar J_\bd$ shows that the optimal transport
occurs when a large part of the filament is crowded, $L_*\simeq 0.8L$,
but the traffic jam is not yet too dense.


\subsection*{Two-state model}


To get some further insight into the properties of these
self-organized density profiles, we studied the stationary states of
these systems using a two-state model. In this model, which is
described in detail in appendix \ref{sec:methods}, the dependence of
the concentration profiles on the radial coordinate is neglected and
motors can be in two states, namely bound and unbound. This
approximation is justified, because the profile of the unbound motor
density depends only weakly on the radial coordinate. The two-state
approximation captures most of the relevant features of these profiles
and numerical solutions for the stationary profiles are obtained much
faster than by Monte Carlo simulations.

For the uniaxial systems, the two-state model as defined in appendix
\ref{sec:methods} is given by
\begin{equation}
  \label{master_eq_axon_1}
  v_\bd \rho_\bd(x)\Big[1- \rho_\bd(x+1)\Big] = \phi D_\ub\Big[\rho_\ub(x+1)-\rho_\ub(x)\Big]
\end{equation}
and
\begin{equation}\label{master_eq_axon_2}
  v_\bd \rho_\bd(x)\Big[1- \rho_\bd(x+1)\Big]- v_\bd \rho_\bd(x-1)\Big[1- \rho_\bd(x)\Big] = \tpi\rho_\ub(x)\Big[1- \rho_\bd(x)\Big] -\tilde\epsilon\rho_\bd(x)\Big[1- \rho_\ub(x)\Big].
\end{equation}
These equations express the balance of bound and unbound currents and
the binding to and unbinding from the filament, respectively, see
appendix \ref{sec:methods}.

For the non-jammed low density region, some analytical results can be
obtained from these equations which are presented in appendix
\ref{sec:app:lowDensity}. In particular, an exponential increase of
the density profile is obtained as $\rho_\bd\approx {\cal N}e^{x/\xi}$
with a length scale $\xi$ as given by \eq{eq:xi_aus2state}.

In order to obtain results for arbitrary densities, we solved the
two-state equations \ref{master_eq_axon_1}--\ref{master_eq_axon_2}
numerically.  Some profiles of the bound motor density as well as the
average current as a function of the number of motors are shown in
\fig{fig:dichteProfileTubeNUM} for a relatively large system with
$L=1000$. While the main features are the same as for the smaller
system discussed above, some additional details can be seen here.  The
current increases linearly with the number $N$ of motors for small
$N$, but at a certain point, $N\simeq 500$ or $N/L\simeq 0.5$ in
\fig{fig:dichteProfileTubeNUM}(b), the slope begins to change. The
current then increases more slowly, but again nearly linearly, until
it reaches its maximum. This change in slope of the current
corresponds to the formation of a plateau in the density profile,
where the density in the traffic jam is approximately constant and
changes only little upon addition of motors.

Until now, we have assumed thermal detachment, i.e., that detachment
at the end of the filament occurs with the same rate as detachment at
any other site of the filament. As mentioned before, see
\eq{thermalunbindlast}, a second possibility is active unbinding,
i.e., that motors detach with an increased rate at the end by making
an active step which leads to unbinding. There is some indirect
evidence for a quicker detachment of kinesin motors at the microtubule
end from experiments and computer simulation of the formation of aster
and vortex patterns of microtubule by motors
\citep{Surrey__Karsenti2001}.  In simulations, quick detachment at the
microtubule end leads to the formation of vortex (or spiral-like)
structures, while slow detachment is necessary for the formation of
asters-like centered arrangements.  Kinesin is able to form both
asters and vortices, suggesting that detachment at the microtubule end
is relatively quick, while the kinesin-related motor ncd only forms
asters and thus probably detaches slowly at the microtubule end
\citep{Surrey__Karsenti2001}.

We have determined density profiles for both cases using the two-state
approach, see \fig{fig:profilemodifmodel}. These density profiles show
that the jamming behavior is rather insensitive to the motor behavior
at the end of the filaments. Except for the region very close to the
filament end, the profiles for the two cases agree well. In
particular, the domain wall or interface represented by the steep
increase of the density profile at the beginning of the traffic jam is
the same in both cases.  This observation shows that the traffic jam
is not due to the slow unbinding at the end, but due to the
accumulation of motors with an exponential density profile, which
follows from the balance of bound drift current and unbound diffusive
currents in a uniaxial geometry. On the other hand, the density
profile within the jammed region depends strongly on the detachment
rate at the filament end. While there is a weak increase of the bound
density inside the jammed region for small detachment rate at the
filament end, the bound density decreases strongly in this region in
the case of an increased detachment rate at the end, see
\fig{fig:profilemodifmodel}.

Comparing the density profiles obtained from the two-state model with
those obtained from simulations for the same parameter set, we find
quite good agreement in the case where the entire filament is crowded
by motors. For smaller overall motor concentrations, qualitative
agreement is still good (except for the region close to the left
boundary, see appendix \ref{sec:app:lowDensity}), but there are small
quantitative discrepancies.  We find that the length scale $\xi$ of
the exponential increase of the density is smaller in the two-state
model than in the simulations.  Correspondingly the crowded region is
slightly longer in this approximation. This difference is due to
neglecting a depletion zone close to the filament in the two-state
model. Close to the filament, the unbound motor density is smaller
than its radial average in the low-density region.  Taking it as
independent of the radial coordinate, we thus overestimate binding of
motors to the filament.  Therefore there are more motors bound to the
filament in the two-state model than in the simulations, which results
in a longer traffic jam, and the maximum of the current is shifted to
a smaller number of motors in the system.

In order to obtain a quantitative description of the radial density
profiles, we have solved the full three-dimensional diffusion equation
for the low-density region and derived an analytical expression for
the depletion layer close to the filament as shown in appendix
\ref{sec:app:depletionLayer}.  It follows from the latter expression
that the radial profile of the unbound density is nearly flat far from
the filament and exhibits a logarithmic depletion zone close to the
filament. This confirms the observation that the unbound motor density
depends only weakly on the
radial coordinate 
which justifies the two-state approach. Comparing the results from
this calculation with the simulation of the full model, good agreement
is found. In \fig{fig:deplLayer}, we have plotted both the
longitudinal (a) and radial profiles (b) as obtained by both methods.
The radial profiles exhibit the predicted depletion layer close to the
filament in the low density region to the left of the traffic jam.  In
the crowded region, the unbound density is enhanced close to the
filament in comparison to the value far from the filament. The full
diffusion equation also leads to a condition for the length scale
$\xi$ given by \eq{bedFuerXi} and we obtain $\xi\simeq 37.4$ for the
parameters used in \fig{fig:deplLayer}, in good agreement with the
value from simulations which is $\xi\simeq 37$.  In contrast, the
two-state approximation yields the smaller value $\xi\simeq 24$,
because it overestimates the current of motors binding to the
filament.


\section*{Density profiles for centered filament systems}
\label{sec:asters}


In this section, we consider profiles of the motor concentration in
centered filament systems or aster-like arrangements of filaments as
shown in \fig{fig:tubeAsterGeom}(c). Such arrangements can be formed
for microtubules in vitro either by nucleation from
microtubule-organizing centers \citep{Holy__Leibler1997} or by
self-organization of microtubules and motor complexes
\citep{Nedelec__Leibler1997,Surrey__Karsenti2001}.  Centered filament
systems mimic the most common organization of microtubules in cells.
Motivated by the restructuring of this organization during cell
division and the formation of the mitotic spindle
\citep{Hyman_Karsenti1996}, many experiments have focused on the case
where the filaments are also mobile.

In the following, we consider immobilized aster-like arrangements of
filaments which are not reorganized by the action of motors.  The
asters consist of $N_f$ filaments of length $L$ arranged radially in a
thin disk of radius $L$ and height $h$. We take the filaments to
extend from $r=0$ to $r=L$ within the disk, but again smaller
filaments lead to very similar results.  In that case, active directed
currents of motors along the filaments and diffusive motor currents
will again be balanced in a stationary state.  For the aster geometry
some theoretical and experimental results for low motor densities have
recently been reported by 
\citet{Nedelec__Maggs2001}.\footnote{In the experiments of
  \citet{Nedelec__Maggs2001} the dynamics is more complicated, since
  they used mobile filaments and their motor constructs can also
  displace these filaments with respect to each other, so that the
  motor density profiles and the filament patterns develop in
  coordination. After some time, however, these systems reach a steady
  state, in which the filament pattern is stationary (although not
  completely immobile) and can, on average, be represented by a fixed
  filament system. In addition, once the aster-like structure is
  formed, the filaments are usually sufficiently separated from each
  other (with the exception of the center of the aster), so that the
  additional dynamics plays only a minor role. Indeed, in the low
  density case, the theoretical density profiles agree well with the
  experimental profiles, as we will discuss below. In order to exclude
  the more compicated dynamics, one could immobilize the filaments
  once the stationary state has been reached or, alternatively, use
  centered microtubule systems nucleated from centrosomes
  \citep{Holy__Leibler1997} and conventional kinesins which cannot
  bind to two filaments at the same time.} We confirm their main
theoretical result, an algebraic density profile far from the center
of the aster, and extend the study of concentration profiles in asters
by exploring the effect of mutual exclusion.

\subsection*{Two-state equations for centered systems}

Centered filament systems are implemented in the two-state model by
substituting the coordinate $n$ used in the general expressions with
the radial coordinate $r$, by using \eqs{j_out}--\ref{j_ub} for the
bound and unbound motor currents and introducing a geometrical weight
factor $\phi(r)\approx\phi_0 r$ as described in appendix
\ref{sec:methods}. The latter factor implements the fact that with
increasing $r$, the volume available for unbound diffusion increases.
The two-state model equations are then given by
\begin{equation}
  \label{master_eq_aster_1}
  v_\bd \rho_\bd(r)\Big[1- \rho_\bd(r+1)\Big] =  D_\ub\, \phi(r) \Big[\rho_\ub(r+1)-\rho_\ub(r)\Big] 
\end{equation}
\begin{equation}\label{master_eq_aster_2}
  v_\bd \rho_\bd(r)\Big[1- \rho_\bd(r+1)\Big]- v_\bd \rho_\bd(r-1)\Big[1- \rho_\bd(r)\Big] = \tpi\rho_\ub(r)\Big[1- \rho_\bd(r)\Big] -\tilde\epsilon\rho_\bd(r)\Big[1- \rho_\ub(r)\Big]
\end{equation}
for the case of motors moving outwards in an aster.  If motion of
motors is directed inwards, i.e.\ if $v_\bd<0$, the two-state
equations are given by
\begin{equation}
  \label{master_eq_aster_in_1}
  v_\bd \rho_\bd(r+1)\Big[1- \rho_\bd(r)\Big] = \phi(r) D_\ub\Big[\rho_\ub(r+1)-\rho_\ub(r)\Big] 
\end{equation}
\begin{equation}\label{master_eq_aster_in_2}
  v_\bd \rho_\bd(r+1)\Big[1- \rho_\bd(r)\Big]- v_\bd \rho_\bd(r)\Big[1- \rho_\bd(r-1)\Big] = \tpi(r)\rho_\ub(r)\Big[1- \rho_\bd(r)\Big] -\tilde\epsilon\rho_\bd(r)\Big[1- \rho_\ub(r)\Big].
\end{equation}

In the low density limit, these equations lead to the algebraic
density profile $\rho_\bd\sim r^{\eta}$ reported by
\citet{Nedelec__Maggs2001}, see appendix \ref{sec:app:lowDensity}. The
exponent $\eta\sim v_\bd$ is positive if motors move outwards and
negative if motors move inwards.

\subsection*{Numerical results}

To study the effect of hard core exclusion in asters, we used the
parameters given by N\'ed\'elec {et al.}\ for the motor constructs
used in their experiments \citep{Nedelec__Maggs2001}.  In the numerics
all parameters are given in units of the microtubule periodicity
$\ell=8\,$nm and the step time $\tau_s=10\,$ms. Parameters of the
bound state are: $v_\bd=0.8\,\mu{\rm m/s}=\ell/\tau_s$ and
$\tilde\epsilon=0.01\,\tau_s^{-1}$ corresponding to unbinding after
100 steps, and those of the unbound state:
$D_\ub=20\,\mu{\rm m}^2/{\rm s}=3125\,\ell^2/\tau_s$ and
$\tpi=2.6\,\mu{\rm m}^2{\rm s}^{-1}/\ell^2=405.6\,\tau_s^{-1}$. 
Parameters which correspond to kinesin with beads as used by
\citet{Lipowsky__Nieuwenhuizen2001} lead to similar results.  All
results shown in the following are obtained for an aster of $N_f=300$
microtubules of length $50\,\mu{\rm m}=6250\,\ell$, which is confined
in a slab of height $9\,\mu{\rm m}=1125\,\ell$.

\subsubsection*{Motors moving inwards}

We consider first the case where motors move inwards. For this case,
experimental results have been reported by 
\citet{Nedelec__Maggs2001}. Accumulation of motors in the center of
the aster is observed by fluorescence microscopy. Profiles of the
total motor concentration, i.e.\ the concentration averaged over bound
and unbound motors,
\begin{equation}
  \bar\rho(r)=\frac{\rho_\bd(r)+\phi(r)\rho_\ub(r)}{1+\phi(r)}\simeq \frac{1}{\phi(r)}\rho_\bd+\rho_\ub\sim \frac{1}{\phi_0}r^{\eta-1}+\frac{\tilde\epsilon}{\tpi} r^\eta,
\end{equation}
can be extracted from the fluorescence images.  The last expression is
valid for small motor densities and sufficiently large values of $r$
and predicts that the density profile
exhibits a crossover from a decay $\sim r^{\eta-1}$ for small $r$ to
$\sim r^\eta$ for large $r$. This crossover behavior is seen in the
experimental data \citep{Nedelec__Maggs2001}.

For small overall motor concentrations, the numerical solution of the
master equations exhibits the power law behavior predicted
theoretically by neglecting exclusion effects. For the chosen
parameters we find $\rho_\ub(r)\sim r^\eta$ with $\eta\simeq -0.54$
from the data for $N=10^4$ shown in \fig{fig:mDichte_InOut}(a) in
agreement with \eq{eta}. In the center of the aster, a traffic jam
is obtained already for small total number of motors. The traffic jam
is however rather short and, in contrast to the case of uniaxially
aligned filaments, does not grow substantially in length, when the
number of motors in the system is increased, see
\fig{fig:mDichte_InOut}(a).  Jamming of motors occurs only for small
$r$ ($\lesssim 20\ell\simeq 0.2\mu$m). For this range of $r$, no
experimental data are available. 
In contrast to the case of uniaxial systems, the motor behavior at the 
end of the filaments is crucial for the presence of jams in centered 
filament systems:
If motors unbind actively at the
filament ends (in the center of the aster), traffic jams are absent as 
shown in \fig{fig:aster_end}(a).
In the regions with low motor densities, the density profiles for thermal 
(slow) and active (fast) detachment at the ends of the filaments agree 
perfectly.

The main effect of the mutual exclusion is that density profiles
get more and more flat with increasing motor concentration in the
system, see \fig{fig:mDichte_InOut}(a). This means that the power law
profile is found only for small overall motor concentrations.  The
average current in the system exhibits again a maximum at an optimal
motor concentration. The maximum occurs at a motor concentration,
where the bound motor density becomes nearly constant and the power
law behavior is hardly identified.

\subsubsection*{Motors moving outwards}

For motors moving outwards in an aster we obtain profiles as shown in
\fig{fig:mDichte_InOut}(b). For small numbers of motors (and not too
close to the boundaries) the bound density follows the power law 
obtained from the linear equations.  Now the exponent $\eta$ is
positive but small.  With increasing motor concentration, the profile
of the bound density gets again more and more flat and the filaments
become more and more crowded. As in the case of outward movements,
however, the jams at the end of the filaments grow only very weakly
and remain rather small ($\lesssim 50\ell\simeq 0.5\mu{\rm m}$). The
motor behavior at the filament ends is crucial for the presence of
these jams also in this case, and jams are absent if motors unbind
quickly at the filament ends, see \fig{fig:aster_end}(b).

The new feature compared to the previous case is that the profile of
the total motor concentration, which is rather flat for small motor
concentration, develops a pronounced maximum in the center of the
aster as the number of motors is increased beyond the optimal motor
concentration, see \fig{fig:mDichte_InOut}(b).  This can be understood
in the following way: If no ATP is added to the system, motors will
accumulate in the center of the aster, simply because they bind
strongly to the filaments, and, in the center, the number of binding
sites per unit area is maximal. If ATP is added, motors are driven
outwards by active directed motion. If now the number of motors in the
system is increased, so that the motor movements are slowed down by
the exclusion effect, the outward drift is suppressed and accumulation
in the center is successively restored.


\section*{Discussion}
\label{sec:discussion}


We have presented theoretical results for the density profiles of
molecular motors in arrays of cytoskeletal filaments. Motors were
described as particles which move actively, i.e.\ in a directed
manner, when they are bound to cytoskeletal filaments, but undergo
non-directed diffusion upon unbinding from filaments. In addition,
motor particles interact via mutual exclusion.  On the one hand, these
models are designed to describe the generic behavior of the movements
of molecular motors; on the other hand, the model parameters can be
adapted to the transport properties of specific motor molecules. In
general, these models involve certain simplifications compared to real
systems.
We have therefore tested a few modifications of the models to check
that a more detailed modeling does not change our conclusions.

Mutual exclusion of motors is obviously enhanced, if the motors carry
large cargoes such as latex beads or vesicles. Furthermore,
microtubules consist of 12--14 protofilaments which correspond to
12--14 parallel tracks \citep[see also][]{Nieuwenhuizen__Lipowsky2002,Nieuwenhuizen__Lipowsky2004}. For the uniaxial geometry, we have performed 
simulations of lattice
models for which these two features have been incorporated.  In these
latter simulations, motors which occupy a cubic volume of $M^3$
lattice sites move on a microtubule consisting of 12 protofilaments
arranged in a tubular geometry. In the simulations we chose $M=3$ and
$M=5$. These cargoes can mimic small vesicles with diameters of some
tens of nanometers. In vivo, the cargo diameters lie between a few
nanometers for a single protein or RNA molecule and hundreds of
nanometers for a large organelle. In addition, these model cargoes
attach to the filament only with one of their surface sites, which
represents the motor. Therefore, those cargoes which are not bound to
the filament have an additional rotational degree of freedom.  The
resulting density profiles are averaged over $M$ subsequent lattice
sites, because the unrealistic cubic shape of the cargoes and the
rigidity of their attachment to the motors leads to an artificial
sublattice structure in the crowded region.  We then obtain density
profiles that resemble the ones discussed above, but the value of the
bound density in the crowded region is smaller, because a smaller
number of motors can block the filaments.  In particular, a motor
bound to one protofilament also blocks binding sites of the adjacent
protofilaments because of the steric hindrance induced by its large
cargo.

Finally, let us relate our results to experiments. We have determined
profiles of the motor density and motor currents in uniaxial and
centered filament systems. On the one hand, these systems are directly
accessible to experiments in biomimetic model systems in vitro.
Density profiles as discussed here have so far only been measured for
the case of centered or aster-like systems \citep{Nedelec__Maggs2001} 
(see our footnote at the beginning of the section on centered filament 
systems for a discussion of the dynamics in these experiments).
The latter experiment shows the power law profile which is obtained
from the theory for low motor densities. Higher motor densities and
the corresponding jamming behavior have not been explored in this
experiment, but could be studied in the same way by increasing the
overall motor concentration. For the latter case, our theoretical
study makes detailed predictions for the density and current profiles
which could be checked in such an experiment.  In addition, it would
be quite interesting to construct other filament arrangements and
compartment shapes and to study the corresponding motor transport
experimentally.

On the other hand, we can also compare our theoretical results about
motor traffic in closed compartments with experimental studies on
motor traffic in biological cells, where, however, additional
phenomena such as the dynamics of the filaments, the regulation of the
motor activity and the presence of other cellular structures play also
important roles.  Using fluorescence probes, several groups have
measured the density profiles of molecular motors in vivo.  One
particular interesting system are kinesin motors in fungal hyphae.
These hyphae are tubular compartments which contain uniaxial filament
systems. In one experiment, strong localization of kinesin has been
observed at the tip of these fungal hyphae
\citep{Seiler__Schliwa2000}. The comet-like density profiles of these
motors localized at the tip correspond to the case of low motor
density in our model.  However, this localization is only found for
kinesin mutants lacking a certain regulatory domain, i.e.\ for motors
which move actively, but which are not regulated by cargo binding
\citep[see also][]{Verhey__Rapoport1998}.  The underlying regulatory
mechanism is the deactivation of the motor via folding of its tail if
no cargo is bound to it \citep{Coy__Howard1999,Seiler__Schliwa2000}.
The deactivated motors do not exhibit active movement along filaments
and can diffuse back over larger distances. Further regulatory
mechanisms have mainly been discussed for the case of axons where the
question, whether and how motors are transported back is most
prominent \citep{Goldstein_Yang2000}. The mechanisms include local
degradation of motors at the axon terminal
\citep{Dahlstroem__Brady1991} and backward transport by motors of
opposite directionality
\citep{Hirokawa__Kawashima1990,Hirokawa__Brady1991}.

Very recently, another fungal kinesin was also shown to localize at
the tip of the hyphae and to exhibit these comet-like profiles. In
this case, larger motor concentrations were induced by increasing the
level of expression of the corresponding gene. This leads to density
profiles with regions of high motor density which increase in length
with increasing expression level
\citep{Konzack2004,Konzack__Fischer2004}. According to our model,
these density profiles should represent growing traffic jams. It would
be highly desirable to repeat these experiments in vitro.

In summary, we have discussed theoretically the stationary density and
current profiles of molecular motors in uniaxial and centered
aster-like arrangements of cytoskeletal filaments. In particular, we
have explored the effects of exclusion and jamming which can be
addressed in these systems by varying the overall motor concentration.
The two types of filament systems, which we studied, exhibit different
density profiles and different jamming behavior.  For small overall
motor concentrations, the profiles are exponential in uniaxial
systems, but algebraic in centered systems except for the crowded
region close to the filament ends.  Increasing the overall motor
concentration, the jammed region grows in the uniaxial geometry,
resulting in the coexistence of large regions of high and low density
of bound motors, while the crowded region remains small in centered
systems, where larger overall motor concentrations lead to a
flattening of the profile if the motors move inwards, and to the
build-up of a concentration maximum in the center of the aster if
motors move outwards. In addition, the jamming in the uniaxial systems
is rather insensitive to the motor behavior at the ends of the
filaments, while the latter behavior is crucial for the presence of
jams in centered systems.

Both geometries studied here mimic arrangements of filaments in cells
and are accessible to in vitro experiments. The predictions for both
geometries can thus be tested experimentally. Some density profiles
have been determined experimentally. These profiles correspond mainly
to the case of low motor density
\citep{Nedelec__Maggs2001,Seiler__Schliwa2000} --- only one recent
experiment \citep{Konzack2004,Konzack__Fischer2004} addresses higher
motor densities --- and are in agreement with our theoretical
description.

\section*{Appendix}
\appendix
\renewcommand{\theequation}{\Alph{section}.\arabic{equation}}
\setcounter{equation}{1}


\section{Theoretical methods}
\label{sec:methods}


\subsection*{Monte Carlo simulations}

We performed Monte Carlo simulations for the case of uniaxial
arrangements of filaments, where filaments are located within a
cylindrical tube-like compartments and aligned parallel to the
cylinder axis which we take to be the $x$-axis.  We take the filaments
to have the same length $L$ as the tube, but we checked that shorter 
filaments lead to
very similar results. The cylindrical tube with radius $R$ is taken to
consist of all 'channels', i.e.\ lines of lattice site parallel to the
filament, with $u\equiv (y^2+z^2)^{1/2}\leq R$ and $0\leq x\leq L$.
Reflecting boundary conditions are implemented by rejecting all moves
to lattice sites outside this range.  Within the closed tube the
number $N$ of motors is fixed. Each Monte Carlo step, corresponding to
a unit of the basic time scale $\tau$, consists of $N$ Monte Carlo moves.
At each move, a motor particle is chosen randomly and updated
according to the random walk probabilities.

\subsection*{Two-state model}

Our Monte Carlo simulations show that the stationary profiles of the
unbound motor density depend only weakly on the coordinates
perpendicular to the filaments. In order to determine the stationary
state, we can therefore use a two-state approximation, in which all
unbound 'channels' are treated as equivalent and the motors can be in
two states, bound and unbound. The stationary state is then
characterized (i) by the balance of bound and unbound currents,
$j_\bd$ and $j_\ub$, respectively, as given by
\begin{equation}
  \label{master_eq_allg_1}
  j_\bd (n)= \phi(n) j_\ub(n)  
\end{equation}
with $0<n<L$ and (ii) by the change of the bound current as a function
of the spatial coordinate $n$ arising from the binding and unbinding
of motors which leads to
\begin{equation}\label{master_eq_allg_2}
  j_\bd(n)- j_\bd(n-1) = \tpi\rho_\ub(n)\Big[1- \rho_\bd(n)\Big] -\tilde\epsilon\rho_\bd(n)\Big[1- \rho_\ub(n)\Big].
\end{equation}
The latter equation expresses the fact that, in the stationary state,
the sum of all outgoing current is equal to the sum of all incoming
currents at any filament site $n$ and corresponds to Kirchhoff's first
rule for electric circuits. Here, $\rho_\bd$ and $\rho_\ub$ are the
local number densities of bound and unbound motors, respectively. (In
the next step, we will express the currents $j_\bd$ and $j_\ub$ in
terms of these densities.) The coordinate $n$ along the filament is
given by the spatial coordinate $x$ along the cylinder axis and the
radial coordinate $r$ for uniaxial and radial arrangements of
filaments, respectively. $\phi(n)$ is a geometrical factor and will be
explained below. The binding and unbinding rates have been rescaled in
\eq{master_eq_allg_2}, $\tilde\epsilon=2\epsilon/3$ and
$\tpi=2\piad/3$.

In addition, we express the bound and unbound motor currents as
functions of the motor densities.  For the tube geometry, we use the
convention that the bound motors move to the right (the case that they
move to the left is then obtained via the reflection symmetry). The
bound motor current is then given by
\begin{equation}\label{j_tube}
  j_\bd(x)=v_\bd \rho_\bd(x)[1- \rho_\bd(x+1)],
\end{equation}
where $v_\bd$ is the velocity in the absence of other motors. In the
presence of many motors, forward steps are only possible if the
filament site in front of a motor is not occupied. The probability of
a vacant site is given by $[1-\rho_\bd]$, which leads to the reduction
of the current as a function of density expressed in \eq{j_tube}.

For radial arrangements of filaments, we have to distinguish inward
and outward movements of bound motors. The bound motor current is
given by
\begin{equation}\label{j_out}
  j_\bd(r)=v_\bd\rho_\bd(r)[1-\rho_\bd(r+1)]
\end{equation}
and
\begin{equation}\label{j_in}
  j_\bd(r)=v_\bd\rho_\bd(r+1)[1- \rho_\bd(r)]
\end{equation}
for outward and inward movements, respectively.
 
In all cases, the diffusive current of unbound motors is given by
\begin{equation}
  \label{j_ub}
  j_\ub(n) = D_\ub\Big[\rho_\ub(n+1)-\rho_\ub(n)\Big]
\end{equation}
with the diffusion coefficient $D_\ub$ of unbound motors. Note that
the latter expression is a discrete version of the usual diffusive
current $D_\ub\partial\rho_\ub/\partial n$.

The geometrical factor $\phi$ introduced in \eq{master_eq_allg_1}
describes the relative weight of the bound and unbound currents and is
given by the number of filament 'channels' per non-filament 'channel'.
In general, $\phi$ is a function of the coordinate $n$.  For $N_f$
isopolar parallel filaments within a cylindrical tube, $\phi$ is given
by $\phi\approx\pi R^2/N_f$. In particular, for a single filament,
$\phi$ is given by the tube cross-section. Notice that, within the
two-state model, the number of filaments appears only via this
geometrical factor and leads to a rescaling of the accessible volume
for the diffusion of unbound motors. In the case of centered filament
systems, the volume available for the unbound diffusion depends on the
radial coordinate $r$, and the geometrical factor $\phi$ increases
linearly with $r$. In this case, $\phi$ is given by the ratio of the
free surface (i.e.\ not covered by filaments) to the area covered by
filament 'channels' which leads to
\begin{equation}\label{def_phi0}
  \phi(r)=\frac{2\pi r h - N_f \ell^2}{N_f \ell^2}\approx\frac{2\pi r h}{N_f \ell^2}\equiv\phi_0 r,
\end{equation}
where $N_f$ is the number of filaments, $\ell^2$ is the cross-section
of a single 'channel', and $h$ is the height of the slab, into which
the aster is confined.

At the boundaries, $x=0$ and $x=L$, terms corresponding to currents
through the tube walls have to be omitted in \eq{master_eq_allg_2}.
Together with the normalization condition
\begin{equation}\label{master_eq_axon_norm}
  \sum_{n=0}^{L}\Big[\rho_\bd(n)+\phi(n)\rho_\ub(n)\Big]=\frac{N}{N_f},
\end{equation}
which fixes the total number $N$ of motors in the tube, these
equations form a system of $2L$ nonlinear equations for the $2L$
unknown densities $\rho_\bd(n)$ and $\rho_\ub(n)$ with $0<n\leq L$.
We have solved this system of non-linear equations numerically using
Newton's method with backtracking \citep{NumericalRecipes}.  The
advantage of the two-state approach over the Monte Carlo simulations
is that it requires less computation time, so that larger systems are
accessible. In addition, simulations take particularly long
computation times, if unbound diffusion is fast compared to bound
movement, $D_\ub/v_\bd\ell\gg 1$, which is the case for cytoskeletal
motors without large cargoes. In this case, the basic time scale
$\tau$ of the simulations is much smaller than the step time
$\tau_s\simeq \tau/(1-\gamma)$ of the bound movements because $\gamma$
is close to one. In contrast, in the two-state approach, the necessary
computation time is independent of the parameter values. In addition,
within the two-state approximation, the computation time for several
filaments (arranged in parallel or in an aster) is the same as for a
single filament.

As we do not distinguish between the different non-filament 'channels'
in the two-state model, we neglect depletion layers close to the
filaments as we discuss in Appendix \ref{sec:app:depletionLayer} in
some detail for the tube geometry.  In addition, a mean field
approximation is implicit in the relations for the bound motor current
as given by \eqs{j_tube}--\ref{j_in}. However, a comparison of the
stationary profiles from the two-state approach with simulation
results obtained for the case of truly equivalent unbound channels,
for which the two-state approximation is exact, shows very good
agreement. We therefore conclude that, in contrast to the open tube
systems discussed by \citet{Klumpp_Lipowsky2003}, the mean field
approximation is quite accurate for the closed systems discussed here.


\section{Low density limit of the two-state model}
\label{sec:app:lowDensity}

Some analytical results can be obtained for the non-jammed low density 
regions both in uniaxial and centered filament systems. For this purpose, 
we consider the continuum version of the two-state equations.

\subsection*{Uniaxial systems}

The continuum two-state equations for uniaxial filament systems are
obtained by expanding \eqs{master_eq_axon_1} and
\ref{master_eq_axon_2} up to second order in the lattice
constant\footnote{We expand \eq{master_eq_axon_1} taken at positions
  $x$ and $x-1$ and average the results to get an non-ambiguous
  result.  This agrees with the result obtained by expanding the
  time-dependent equations. The expansion leads to
  $D_\bd=v_\bd\ell/2$, but within the continuum equations, we can also
  treat $D_\bd$ as an independent parameter.} which leads to
\begin{equation}
  v_\bd \rho_\bd (1-\rho_\bd)  - D_\bd \frac{\partial \rho_\bd}{\partial x}=   D_\ub \phi \frac{\partial \rho_\ub}{\partial x} \label{2zust_stat1}
\end{equation}
\begin{equation}
  v_\bd\frac{\partial }{\partial x} \rho_\bd(1-\rho_\bd)-D_\bd\frac{\partial^2\rho_\bd}{\partial x^2} =  \tpi \rho_\ub(1-\rho_\bd)-\tilde\epsilon \rho_\bd(1-\rho_\ub)  \label{2zust_stat2}
\end{equation}
with the boundary conditions $j_\bd=v_\bd \rho_\bd
(1-\rho_\bd)-D_\bd\partial \rho_\bd/\partial x=0$ at $x=0$ and $x=L$
which express the fact that no motors enter or leave the tube. These
boundary conditions imply, via \eq{2zust_stat1}, that also the unbound
motor currents vanish at the boundaries.  In the low-density limit,
hard core repulsion or exclusion can be neglected.  This is
appropriate in the non-crowded region, where $\rho_\bd\ll 1$. For
simplicity, we also neglect the bound diffusion terms, i.e., we
consider the case $D_\bd=0$. On the one hand, this can be understood
as taking into account only the first non-vanishing terms in the
derivation of the continuum equations. On the other hand, a comparison
of numerical solutions of the continuum equations with and without
this terms shows, that the precise value of $D_\bd$ is largely
irrelevant for the solution, as long as the detachment rate is small,
which, however, is the case for processive motors. In the low-density
limit the equations become linear,
\begin{eqnarray}
  v_\bd \rho_\bd & = &  D_\ub \phi \frac{\partial \rho_\ub}{\partial x} \\
  v_\bd\frac{\partial \rho_\bd}{\partial x} & = & \tpi \rho_\ub-\tilde\epsilon \rho_\bd,
\end{eqnarray} 
and, in general, the solution is given by a sum of two exponential
terms. One term, however, decreases exponentially with $x$ and
therefore contributes only close to the left boundary, where it
ensures the boundary condition of vanishing current and leads to a
larger initial slope of the density profile. For sufficiently large
$x$, the solution is therefore increasing exponentially along the
tube,
\begin{equation}
  \rho_\bd(x)\approx{\cal N}  e^{x/\xi},
\end{equation}
where ${\cal N}$ is a constant and
\begin{equation}\label{eq:xi_aus2state}
  \xi=\frac{2v_\bd/\tilde\epsilon}{\left(1+\frac{4\tpi}{D_\ub\phi}\frac{v_\bd^2}{\tilde\epsilon^2}\right)^{1/2} -1}\approx
\frac{\tilde\epsilon D_\ub\phi}{\tpi v_\bd}.
\end{equation}
The last approximation is valid for small $v_\bd$ and is also obtained
from our first approximation, \eq{firstApprox} above, where we assumed
that unbinding and rebinding are equilibrated.  The unbound density is
given by
\begin{equation}
  \rho_\ub(x)=\frac{\tilde\epsilon}{\tpi} \rho_\bd(x) +\frac{v_\bd}{\tpi} \frac{\partial \rho_\bd(x)}{\partial x}={\cal N}\left(\frac{\tilde\epsilon}{\tpi}+\frac{v_\bd}{\tpi \xi}\right)e^{x/\xi},
\end{equation}
i.e., bound and unbound density are proportional in the low-density
limit. The first term of the factor relating bound and unbound density
is the one obtained in the case of equilibrated transitions between
the bound and unbound states (e.g., from linearizing
\eq{equilibrium_withExclusion}), the second one is a correction which
shows that binding and unbinding are also driven out of equilibrium if
$v_\bd\neq 0$. (Note, however, that this term is of order $v_\bd^2$,
since $\xi\sim 1/v_\bd$, so that up to linear order in $v_\bd$ radial
equilibrium still holds.) This correction term is positive, thus the
current of motors binding to the filament at a given site, $\tpi
\rho_\ub(x)$, is larger than the current of unbinding motors at the
same site, $\tilde\epsilon \rho_\bd(x)$, which is easy to understand,
since the motors bound to the filament are driven away by the drift
$v_\bd$. For small driving velocity $v_\bd$, we can replace the local
balance of binding and unbinding currents at a site $x$ by the
condition
\begin{equation}\label{balance_x_x+v/eps}
  \tilde\epsilon \rho_\bd(x+v_\bd/\tilde\epsilon)\approx\tpi \rho_\ub(x),
\end{equation}
which states that motors binding to the filament at site $x$, move for
a distance $v_\bd/\tilde\epsilon$ before they unbind at site
$x+v_\bd/\tilde\epsilon$. Inserting the solution given above, we can
check that this is fulfilled for small $v_\bd/\tilde\epsilon$.
\begin{equation}
  \tilde\epsilon \rho_\bd(x+v_\bd/\tilde\epsilon)= \tilde\epsilon\,  e^{v_\bd/(\tilde\epsilon \xi)} \rho_\bd(x)\approx (\tilde\epsilon+v_\bd/\xi)\rho_\bd(x)=\tpi \rho_\ub(x).
\end{equation}

The fact that more motors attach to the filament than detach from it,
indicates that this solution cannot be correct for all $x$. In a
system {\it without mutual exclusion}, unbinding will be larger than
binding to the filament only at the end of the filament.  In that
case, we can account for unbinding at the filament end by assuming
that all motors that would have detached in the interval
$[L,L+v_\bd/\epsilon]$ are forced by the boundary to wait at the last
binding site of the filament until they detach.  Therefore the density
at the filament end, $\rho_\bd(x=L)$, is given by
\begin{equation}
  \rho_\bd(x=L)\simeq\frac{1}{\ell}\int_{L}^{L+\frac{v_\bd}{\epsilon}}\d x\, {\cal N}e^{x/\xi}=\frac{\xi}{\ell}\, {\cal N}e^{L/\xi}\left(e^{v_\bd/(\epsilon \xi)} -1\right)\approx \frac{v_\bd}{\epsilon\ell}\, {\cal N}e^{L/\xi},
\end{equation}
where $\ell$ is again the size of the binding site and the last
relation is valid for small velocity $v_\bd$, for which the ansatz
given in \eq{balance_x_x+v/eps} is justified. A comparison with
simulations for the case without mutual exclusion shows good agreement
of the density at the last lattice site with the values obtained by
this procedure.  In reality, however, there is hard core exclusion and
the present solution holds only as long as the bound density is
sufficiently small and breaks down at a certain $x$ because of the
exponential increase of the bound density.

\subsection*{Centered systems}

For centered filament systems, the continuum limit of the two-state
equations for low motor densities leads to
\begin{equation}
  v_\bd\rho_\bd  = \phi_0 r D_\ub\,  \frac{\partial\rho_\ub}{\partial r} +D_\bd \frac{\partial\rho_\bd}{\partial r}
\end{equation}
\begin{equation}
 v_\bd\frac{\partial\rho_\bd}{\partial r} -D_\bd\frac{\partial^2 \rho_\bd}{\partial r^2} = \tpi\rho_\ub-\tilde\epsilon\rho_\bd
\end{equation}
for both inward and outward movements.  In the case $D_\bd=0$, these
equations are equivalent to those used by \citet{Nedelec__Maggs2001}
to describe their experimental results.
These equations lead to
\begin{equation}
  v_\bd\rho_\bd= D_\ub \phi_0\, r \left(\frac{\tilde\epsilon}{\tpi}\frac{\partial\rho_\bd}{\partial r}+\frac{v_\bd}{\tpi}\frac{\partial^2 \rho_\bd}{\partial r^2} -\frac{D_\bd}{\tpi}\frac{\partial^3 \rho_\bd}{\partial r^3}\right) +D_\bd \frac{\partial\rho_\bd}{\partial r}.
\end{equation}
To recover the asymptotic solution of \citet{Nedelec__Maggs2001} we
assume $\rho_\bd\sim r^\eta$ and neglect terms of order $r^{\eta-1}$.
We obtain
\begin{equation}\label{eta}
  \eta=\tpi v_\bd/(\tilde\epsilon\phi_0 D_\ub)
\end{equation}
which can be larger or smaller than zero depending on the sign of the
velocity $v_\bd$.  Note that the bound diffusion coefficient does not
contribute to this asymptotic result.  Interestingly, neglecting terms
of order $r^{\eta-1}$ is equivalent to the assumption that binding to
and unbinding from the filament are balanced locally. Hence
asymptotically, bound and unbound densities are related by
$\rho_\ub=(\tilde\epsilon/\tpi) \rho_\bd$, in contrast to the case of
uniaxial systems, and $\rho_\ub$ decays with the same power law as
$\rho_\bd$.


\section{Depletion layer}
\label{sec:app:depletionLayer}

In this appendix, we derive an analytical expression for the radial
profile of the unbound motor density for the case of a single filament
located along the symmetry axis of a cylindrical tube.  We consider
the linearized diffusion equations which are appropriate for the low
density limit or the non-crowded region to the left of the 'traffic
jam'.

The balance of bound and unbound currents is given by
\begin{equation}\label{balanceBdUb_DeplLay}
  v_\bd\rho_\bd(x)=D_\bd\frac{\partial\rho_\bd(x)}{\partial x}+ D_\ub\, 2\pi \int_\ell^R \d u\, u\, \frac{\partial}{\partial x}\rho_\ub(x,u)
\end{equation}
and the unbound motor density fulfills the stationary diffusion
equation with cylindrical symmetry
\begin{equation}\label{diff_gl1_3d_stat}
  D_\ub\left( \frac{\partial^2}{\partial x^2} +\frac{\partial^2}{\partial u^2}+\frac{1}{u} \frac{\partial}{\partial u}\right)\rho_\ub =0
\end{equation}
which holds for values of the radial coordinate $u$ with $\ell\leq
u\leq R$ with the filament radius $R_{\rm F}\simeq\ell$ and the tube 
radius $R$. The
solution has to fulfill the boundary condition $\partial
\rho_\ub/\partial u=0$ at $u=R$. The longitudinal boundary conditions 
are the same as in the two-state model. Binding to and unbinding from the
filament are described by
\begin{equation}\label{diff_gl2_3d_stat}
  v_\bd\frac{\partial\rho_\bd}{\partial x}-D_\bd\frac{\partial^2\rho_\bd}{\partial x^2} =-\tilde\epsilon\rho_\bd+\tpi \frac{\pi \ell^2}{4}\rho_{\ub}(x,u=\ell),
\end{equation}
which represents the boundary condition for $\rho_\ub$ at $u=\ell$.
The separation ansatz
\begin{equation} \label{factor_ansatz}
  \rho_\ub(x,u)=e^{x/\xi}f(u)\qquad {\rm and}\qquad \rho_\bd(x)={\cal N}e^{x/\xi},
\end{equation}
where ${\cal N}$ is a constant, leads to
\begin{equation}\label{f(r)=2}
  f(u)= \frac{4 f_0}{\pi\ell^2}\, \frac{J_0(u/\xi)Y_1(R/\xi)-J_1(R/\xi)Y_0(u/\xi)}{J_0(\ell/\xi)Y_1(R/\xi)-J_1(R/\xi)Y_0(\ell/\xi)},
\end{equation}
where $J_0$ and $Y_0$ are Bessel functions of order zero of the first
and second kind, respectively, and $J_1$ and $Y_1$ are the
corresponding Bessel functions of the first order \citep{Abramowitz}.
For small $u$ the radial profile 
behaves as $f(u)\sim\ln(u/\xi)$.

The balance of bound and unbound currents, \eq{balanceBdUb_DeplLay},
yields the condition
\begin{equation}\label{bedFuerXi}
  v_\bd=\frac{D_\bd}{\xi}+ D_\ub \frac{4\xi^2}{\pi\ell^2}I(\ell/\xi,R/\xi)\left[\frac{\tilde\epsilon}{\tpi\xi}+\frac{v_\bd}{\tpi\xi^2}-\frac{D_\bd}{\tpi\xi^3}\right]
\end{equation}
with
\begin{equation}
  I(\ell/\xi,R/\xi)\equiv\frac{2\pi}{\xi^2}\int_\ell^R \d u\, u \frac{f(u)}{4/(\pi\ell^2)}=2\pi \int_{\ell/\xi}^{R/\xi} \d z\, z \frac{J_0(z)Y_1(R/\xi)-J_1(R/\xi)Y_0(z)}{J_0(\ell/\xi)Y_1(R/\xi)-J_1(R/\xi)Y_0(\ell/\xi)},
\end{equation}
from which the localization length $\xi$ is determined numerically.

\section{List of symbols}

\begin{tabbing}
  symbol\hspace{1.5cm}\= definition\= \kill

  $D_\bd$  \> (one-dimensional) diffusion coefficient of bound motors \\ 
  $D_\ub$  \> diffusion coefficient of unbound motors \\

  $f(r)$ \> radial part of the concentration profile in a tube \\
  $h$ \> height of the slab to which a filament aster is confined\\
  $j_\bd$ \> local current of bound motors\\
  $j_\ub$ \> local current of unbound motors\\
  $\bar J_\bd$ \> spatially averaged current of bound motors in a closed tube\\
  $\ell$ \> lattice constant, given by the filament repeat distance\\
  $L$ \> linear extension of the compartment,\\
    \> i.e.\ length of the tube or radius of the disk\\
  $L_*$\> length of crowded region (traffic jam)\\
  $n$ \> coordinate along the filament in the general case,\\
      \> $n=x$ and $n=r$ for uniaxial and radial filament arrangements, respectively\\ 
  $N$ \> number of motors \\
  $N_f$ \> number of filaments\\
  $\cal N$ \> normalization constant\\ 
  $r$ \> radial coordinate in the aster geometry\\
  $R$ \> radius of closed tube \\
  $t$ \> time variable\\
  $u$ \> radial coordinate in the tube geometry\\
  $v_\bd$  \> velocity of bound motor \\
  $x$ \> spatial coordinate parallel to the filament\\
  $y$, $z$\> spatial coordinates perpendicular to the filament \\
  
\end{tabbing}

\begin{tabbing}
  symbol\hspace{1.5cm}\= definition\= \kill

  $\alpha$ \> probability for a forward step of a bound motor per unit time $\tau$\\
  $\beta$ \> probability for a backward step of a bound motor per unit time $\tau$\\
  $\gamma$ \> dwell probability of a bound motor per unit time $\tau$ \\
  $\epsilon$\> detachment parameter, $\epsilon/6$ is the detachment probability per non-filament\\
   \> neighbor site per unit time $\tau$\\

  $\tilde\epsilon$\> rescaled detachment probability, $\tilde\epsilon=2\epsilon/3$ \\
  $\eta$ \> exponent of the asymptotic density profiles in asters\\
  $\xi$ \> localization length or decay length of the density profiles\\
  $\piad$ \> sticking probability for a motor hopping to the filament\\
  $\tpi$ \> rescaled sticking probability, $\tpi=2\piad/3$ \\

  $\bar\rho$ \> average motor density (bound and unbound motors)\\
  $\rho_\bd$ \>  density of motors bound to the filament\\
  $\rho_\ub$ \>  density of unbound motors\\

  $\tau$ \> basic time unit, defined by $\tau=\ell^2/D_\ub$\\
  $\tau_s$ \> step time\\
  $\phi$ \> ratio of the number of filament channels to the number of 
  non-filament \\
  \> channels; uniaxial arrangements: tube cross-section/filament number,\\
  \> asters: radius-dependent effective cross-section/filament number\\
  $\phi_0$\> parameter for the ratio of bound to unbound channels in the case of asters\\
\end{tabbing}

\section*{Acknowledgments}

S.K. and R.L. acknowledge support by the Human Frontier Science
Project via research grant RGP 72/2003.  Th.M. N. is grateful for
hospitality at the Max Planck Institute in Golm.

\newpage
\section*{Figure legends}

\noindent
Figure 1:\\
(a) Molecular motors perform active directed movements characterized
by the bound-state velocity $v_\bd$ along a cytoskeletal filament.
After unbinding from the filament, the motor undergoes non-directed
Brownian motion with diffusion coefficient $D_\ub$. As motors are
strongly attracted to filaments, mutual exclusion of motors from
binding sites leads to molecular traffic jams. We study stationary
states for two geometries: (b) uniaxial arrangements of filaments in
closed tube-like compartments and (c) radial or aster-like
arrangements of filaments in closed disk-like compartments.\\[5mm]

\noindent
Figure 2:\\
Profiles of (a) the bound motor density $\rho_\bd$ and (b) the
corresponding bound motor current $j_\bd$ as functions of the
coordinate $x$ along the filament in the closed tube for three
different motor numbers $N$ as obtained from Monte Carlo simulations.
The tube has length $L=200\ell$ and radius $R=25\ell$. The transport
parameters are $\alpha=0.01-2\epsilon/3\simeq 9.93\times 10^{-3}$,
$\beta=0$, $\gamma=0.99$, $\epsilon=10^{-4}$, and $\piad=1$.\\[5mm]

\begin{figure}[h]
  \centering
  \includegraphics[angle=-90,width=\textwidth]{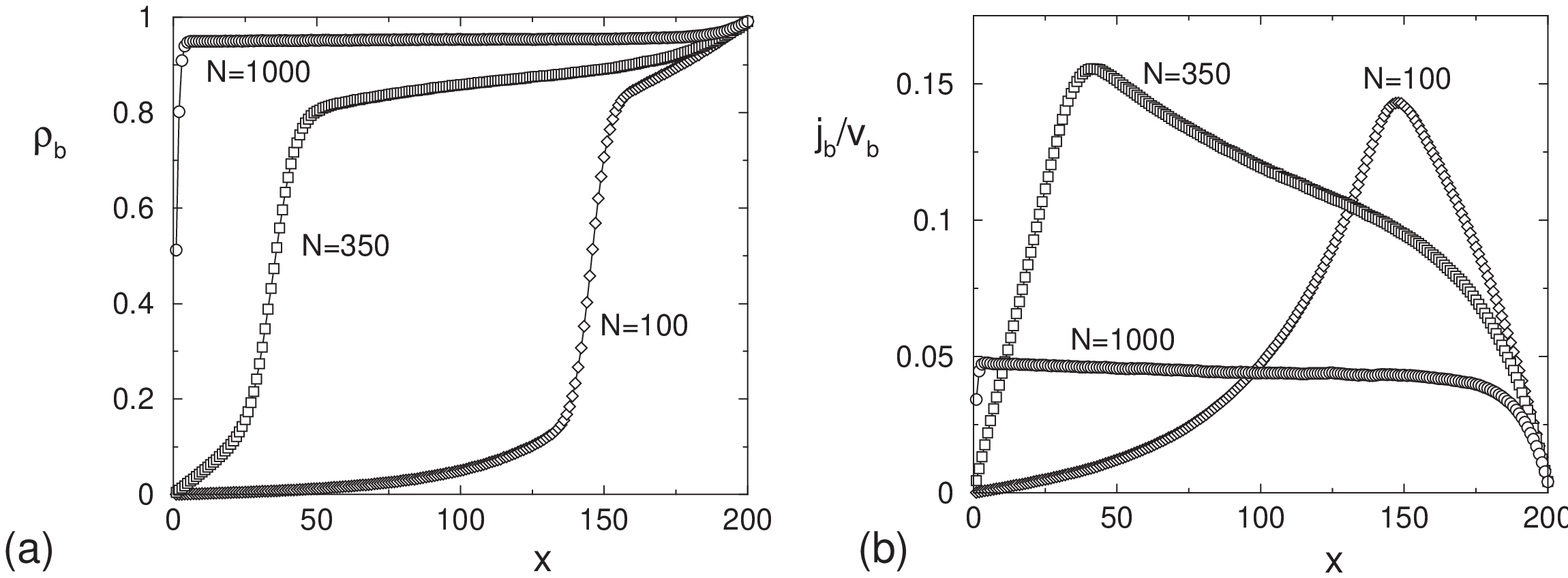}
  \vspace{5cm}\caption{ }
  \label{fig:dichteProfileTube}
\end{figure}

\begin{figure}[h]
  \centering
  \includegraphics[angle=0,width=.6\textwidth]{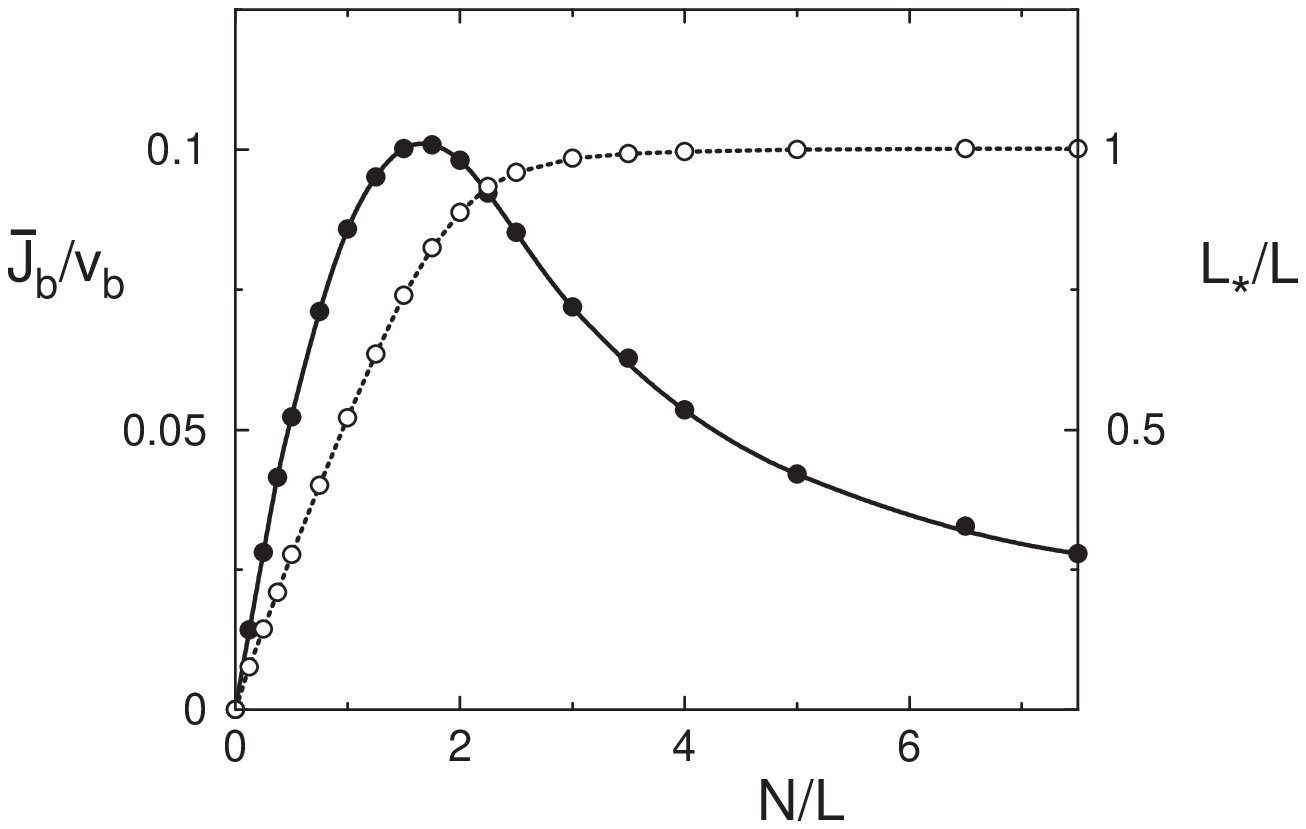}
  \vspace{5cm}\caption{}
  \label{fig:stromMax}
\end{figure}

\noindent
Figure 3:\\
Average current $\bar J_\bd$ of bound motors (filled circles) and
traffic jam length $L_*$ (open circles) as functions of the total
number $N$ of motors in the tube. Geometry and parameters of motion
are the same as in \fig{fig:dichteProfileTube}. The data points at 
$N/L=0.5$,1.74, and 5 correspond to the profile shown in 
\fig{fig:dichteProfileTube}. \\[5mm]

\noindent
Figure 4:\\
Two-state model: (a) Profiles of the bound motor density $\rho_\bd$ as
a function of the spatial coordinate $x$ parallel to the filament and
(b) average bound motor current $\bar J_\bd$ as a function of the
number $N$ of motors in the tube as obtained from the numerical
solution of the discrete two-state model. The chosen tube has length
$L=1000$ and radius $R=25$. The parameters of motion are $v_\bd=0.01$,
$D_\ub=1/6$, $\epsilon=10^{-4}$, $\piad=1$. The numbers of motors in
(a) are (from right to left) $N=100, 200, 300, 400, 500, 1000, 1500,
2000, 2500, 3000, 3500$, and 4000. \\[5mm]

\begin{figure}[h]
  \centering
  \includegraphics[angle=0,width=\textwidth]{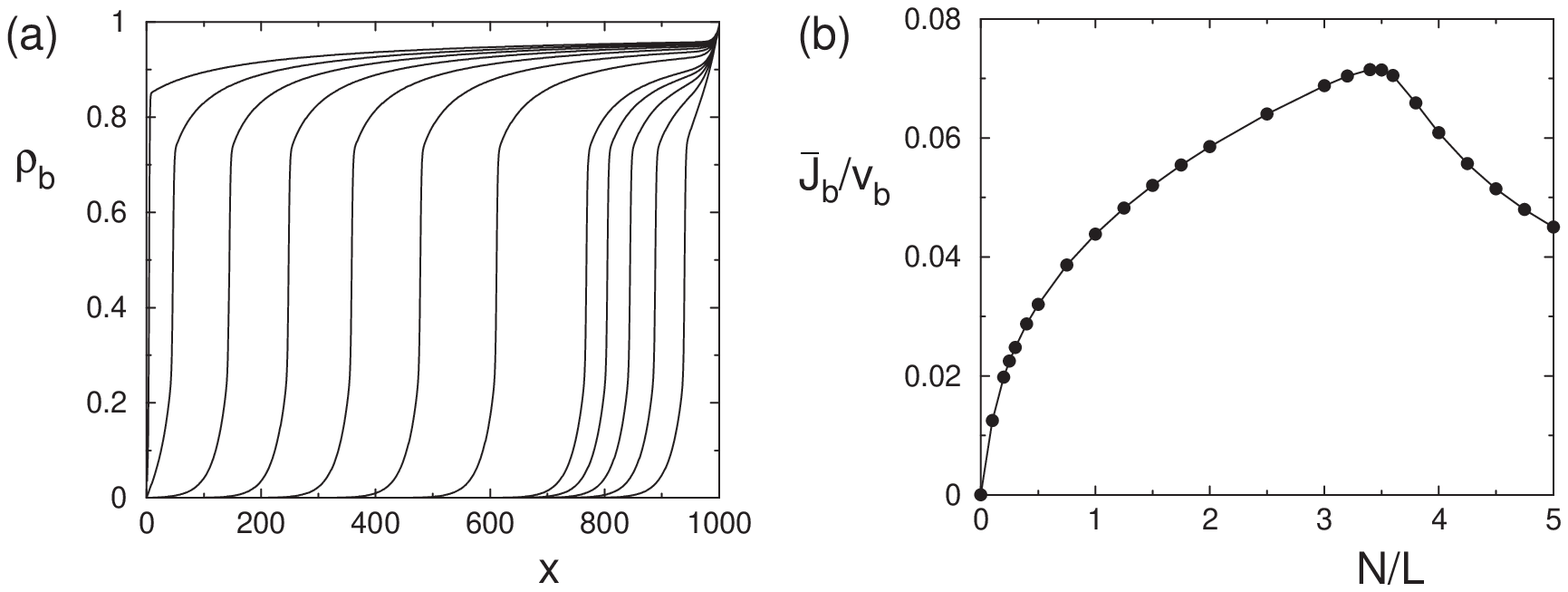}
  \vspace{5cm}\caption{}
  \label{fig:dichteProfileTubeNUM}
\end{figure}

\begin{figure}[h]
  \begin{center}
    \leavevmode
    \includegraphics[angle=0,width=.5\textwidth]{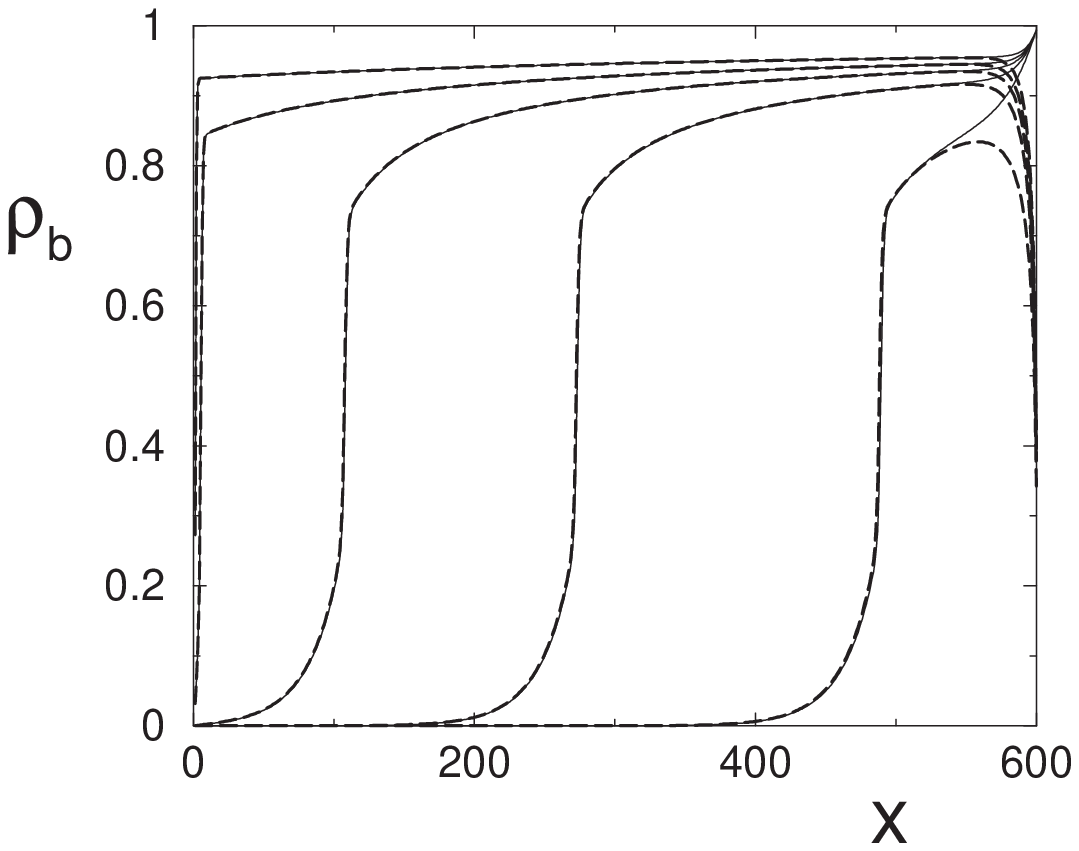}
    \vspace{5cm}\caption{}
    \label{fig:profilemodifmodel}
  \end{center}
\end{figure}

\begin{figure}[h]
  \begin{center}
    \leavevmode
    \includegraphics[angle=0,width=.9\textwidth]{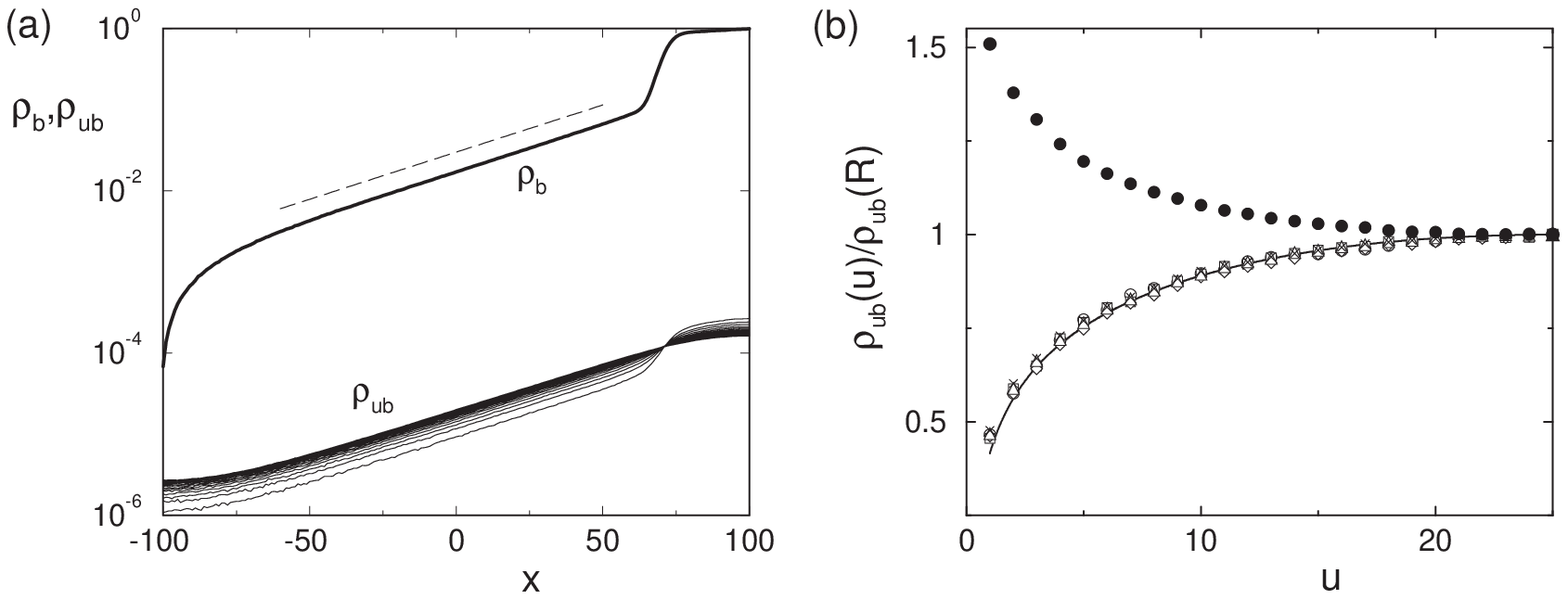}
    \vspace{5cm}\caption{ }
    \label{fig:deplLayer}
  \end{center}
\end{figure}

\noindent
Figure 5:\\
Two-state model: Profiles of the bound motor density $\rho_\bd$ for
active and thermal unbinding of motors at the filament end: A motor at
the end of the filament detaches with the same probability as at any
other filament site (thermal unbinding, solid lines) or by a forward step, 
i.e.\ with
rate $\alpha$ (active unbinding, dashed lines).  $L=600$, $R=25$, parameters of motion
as in \fig{fig:dichteProfileTubeNUM}, and $N=200,800,1400,2000,2600$
(from right to left).\\[5mm]

\noindent
Figure 6:\\
(a) Profiles of the bound motor density $\rho_\bd$ (thick line) and
the radius-dependent unbound motor density $\rho_\ub$ as functions of
the spatial coordinate $x$ parallel to the filament. The lines for the unbound density show the profile at different distances from the filament ($u=1$ to $u=25$, bottom to top in the part left of the jam and top to bottom right of the jam). The dashed line
indicates the exponential $\sim {\rm exp}(x/37.4)$ as obtained from
the linearized equations
\ref{balanceBdUb_DeplLay}--\ref{diff_gl2_3d_stat}. (b) Radial profile
of the unbound density in the low density region, $x=-40
(\circ),-20(\Box),0(\diamond),20(\triangle),40(\times)$, and in the
crowded region, $x=80(\bullet)$. The solid line shows the analytical
result given in \eq{f(r)=2}. The simulation data at all positions in the low density region agree with each other and with the analytical result, which shows that the profile has the product form (\ref{factor_ansatz}). The transport parameters are as in
\fig{fig:dichteProfileTube}, and the tube has length $L=201$ and
radius $R=25$.\\[5mm]

\noindent
Figure 7:\\
Concentration profiles for motors moving (a) inwards and (b) outwards
in aster-like arrays of filaments as functions of the radial
coordinate $r$. Parameters are for motor complexes
as described by 
\citet{Nedelec__Maggs2001}, see text. The numbers of motors are
(from bottom to top) $N=10^4$ ($\circ$), $10^5$ ($\diamond$),
$10^6$ ($\triangle$), $10^7$ ($\Box$), $10^8$~($\star$), $10^9$
($\rhd$).  The profiles shown are profiles of the total motor
concentration $\bar\rho\simeq\rho_\bd/\phi(r)+\rho_\ub$.
Because of the logarithmic scale, discrete data points are only
indicated for small $r$.\\[5mm]

\begin{figure}[h]
   \begin{center}
     \leavevmode
     \includegraphics[angle=0,width=.9\textwidth]{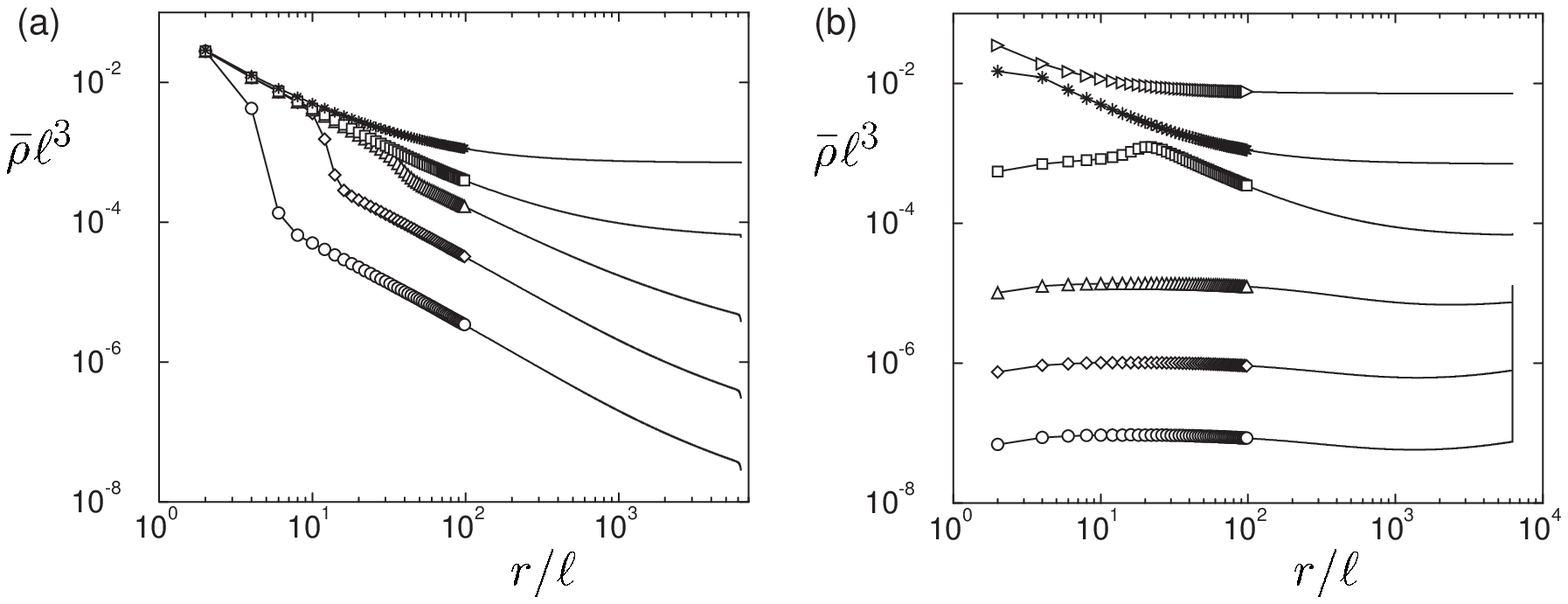}
     \vspace{5cm}
     \caption{ }
     \label{fig:mDichte_InOut}
   \end{center}
\end{figure}

\begin{figure}[h]
   \begin{center}
     \leavevmode
     \includegraphics[angle=0,width=.9\textwidth]{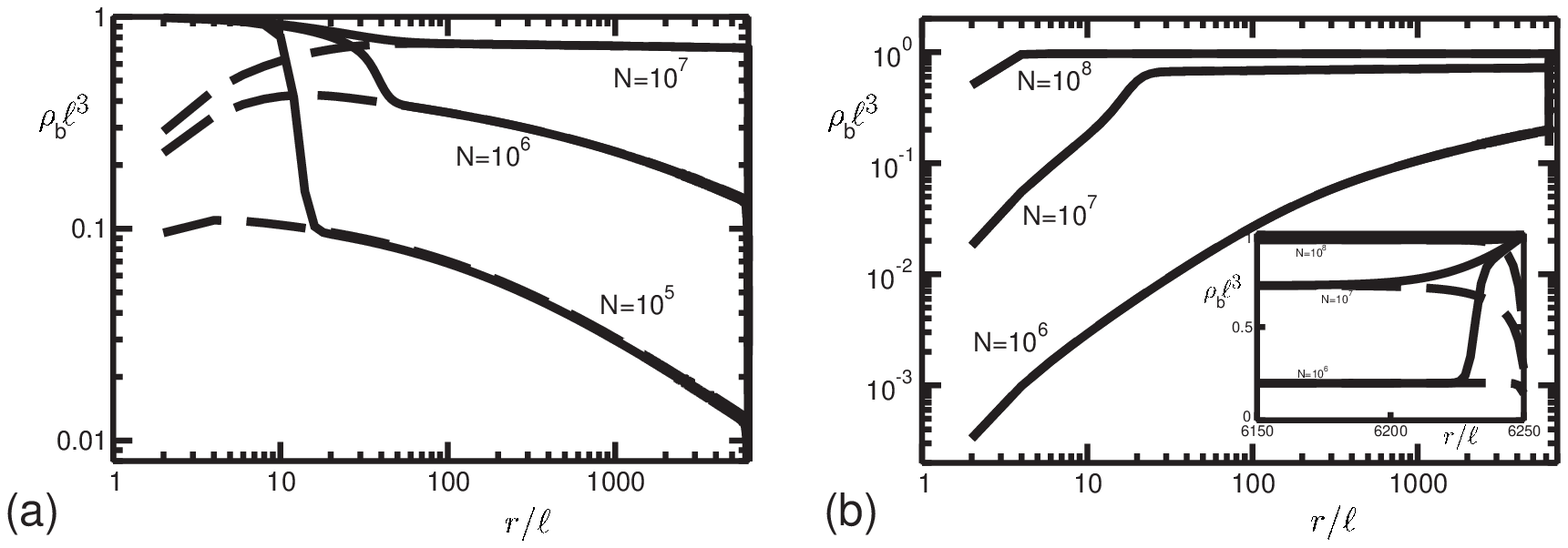}
     \vspace{5cm}
     \caption{ }
     \label{fig:aster_end}
   \end{center}
\end{figure}

\noindent
Figure 8:\\
Thermal versus active unbinding of motors at the filament end in
centered systems: Profiles of the bound motor density $\rho_\bd$ as a
function of the radial coordinate $r$ for motors moving (a) inwards
and (b) outwards in aster-like filament arrays. At the end of the
filaments, motors detach with the same probability as at any other
filament site (solid lines) or by a forward step, i.e.\ with rate
$\alpha$ (dashed lines). The inset in (b) shows the region close to
the filament ends where the profiles for the two cases differ. The
parameters are as in \fig{fig:mDichte_InOut}.

\newpage

\newpage

\newpage

\newpage

\newpage

\newpage

\newpage

\newpage

\end{document}